\documentclass[jkps,preprint,fleqn,showpacs,showkeys]{revtex4}
\usepackage{graphicx}
\usepackage{amssymb}
\usepackage{amsmath}
\usepackage{bm}
\usepackage{color}
\usepackage{ulem}

\begin{document}
\setcounter{page}{0}
\title[]{Ground-state phase diagram of the Kondo lattice model on triangular-to-kagome lattices}
\author{Yutaka \surname{Akagi}}
\author{Yukitoshi \surname{Motome}}
\email{akagi@aion.t.u-tokyo.ac.jp}
\thanks{Fax: +81-3-5841-6817}
\affiliation{Department of Applied Physics, University of Tokyo, Tokyo 113-8656, Japan}

\begin{abstract}
We investigate the ground-state phase diagram of the Kondo lattice model 
with classical localized spins on triangular-to-kagome lattices 
by using a variational calculation.
We identify the parameter regions where a four-sublattice noncoplanar order is stable 
with a finite spin scalar chirality while changing the lattice structure 
from triangular to kagome continuously. 
Although the noncoplanar spin states appear in a wide range of parameters, 
the spin configurations on the kagome network become coplanar 
as approaching the kagome lattice; 
eventually, the scalar chirality vanishes for the kagome lattice model. 
\end{abstract}



\maketitle

\section{INTRODUCTION}
Recently, unusual magnetic orderings have been of intensive research interest 
in geometrically-frustrated spin-charge coupled systems. 
In particular, a spin scalar chiral ordering 
has attracted considerable attention. 
A typical example was theoretically discussed for the Kondo lattice model on a kagome lattice~\cite{Ohgushi_2000}. 
It was shown that a noncoplanar spin ordering 
with the $\mathbf{q}=\mathbf{0}$ three-sublattice structure induces 
an unconventional anomalous Hall effect through the Berry phase mechanism. 
Another example has been recently attracting attention --- 
a scalar chiral ordering with a four-sublattice noncoplanar spin configuration 
on a triangular lattice~\cite{Martin_2008,Akagi_2010,Kumar_Brink_2010,Kato_2010,Akagi_2012}.
The particular order appears in two regions, near 3/4 filling and 1/4 filling~\cite{Akagi_2010}.  
While the former was deduced from the perfect nesting of the Fermi surface~\cite{Martin_2008}, 
the latter is unexpected from the nesting scenario; instead, it was clarified that 
the 1/4-filling chiral state is induced by 
a critical enhancement of effective positive biquadratic interactions 
through the generalized Kohn anomaly~\cite{Akagi_2012}.

The four-sublattice noncoplanar order on the triangular lattice can be viewed as 
an extension of the three-sublattice one on the kagome lattice. 
The kagome lattice is obtained by depleting 1/4 sites periodically in the triangular lattice. 
When omitting 1/4 sites periodically from the four-sublattice order on the triangular lattice, 
one ends up with the $\mathbf{q}=\mathbf{0}$ three-sublattice order with a particular solid angle of three spins on the kagome lattice. 
The stability of the four-sublattice order was studied in detail on the triangular lattice~\cite{Akagi_2010}, 
but that of the three-sublattice order was not investigated on the kagome lattice; 
the spin pattern was set by hand as an internal field for itinerant electrons~\cite{Ohgushi_2000}. 

In this contribution, we investigate the stability of noncoplanar spin ordering 
in the Kondo lattice model on the kagome lattice 
from the viewpoint of connection to the four-sublattice order on the triangular lattice. 
We consider a continuous change of the lattice structure from triangular to kagome 
by modulating the electron hopping between the 1/4 sites and the neighboring sites. 
By using a variational calculation for the ground state 
similar to the previous studies~\cite{Akagi_2010,Akagi_2011Preprint}, 
we clarify the parameter regions where the noncoplanar chiral state becomes stable.

\section{Model and method}
\begin{figure}
\includegraphics[width=5.0cm]{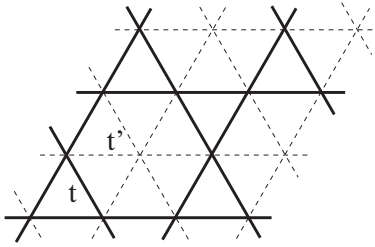}
\caption{
Schematic picture of the triangular-to-kagome lattice. 
The transfer integrals are periodically modulated 
by changing the ratio $t'/t$ 
so that the network connected by $t$ forms the kagome lattice at $t'=0$. 
}\label{fig1}
\end{figure}

We consider the Kondo lattice model on a triangular lattice 
with a modulation of electron hopping to connect the triangular and kagome lattice structures, 
as shown in Fig.~\ref{fig1}. 
We call this the triangular-to-kagome lattice hereafter. 
The Hamiltonian is given by 
\begin{equation}
{\cal H}=- \!\! \sum_{\langle i,j \rangle,\alpha } \!\! t_{ij} ( c^{\dagger}_{i,\alpha } c_{j,\alpha }+\mathrm{h.c.}) 
-J_{\rm H} \! \sum_{i,\alpha ,\beta } \!
    c^{\dagger}_{i,\alpha } \boldsymbol{\sigma}_{\alpha \beta }  c_{i,\beta } \cdot \mathbf{S}_i,  
\label{eq:H}
\end{equation}
where $c^{\dagger}_{i,\alpha}$ ($c_{i,\alpha}$) is a creation (annihilation) operator of conduction electron at
site $i$ with spin $\alpha$, $\boldsymbol{\sigma}$ is the Pauli matrix, 
and ${\mathbf S}_i$ is a localized moment at site $i$. 
The hopping matrix $t_{ij}$ takes $t(=1)$ or $t'$ $(0 \le t' \le 1)$ as being 
the triangular-to-kagome lattice; 
$t'=t$ ($t'=0$) corresponds to the triangular (kagome) lattice. 
The sum in the first term is taken 
for the nearest-neighbor sites on the triangular-to-kagome lattice. 
The sign of the spin-charge coupling $J_{\rm H}$ does not matter, 
since we treat ${\mathbf S}_i$ as a classical spin.

\begin{figure}
\includegraphics[width=8.0cm]{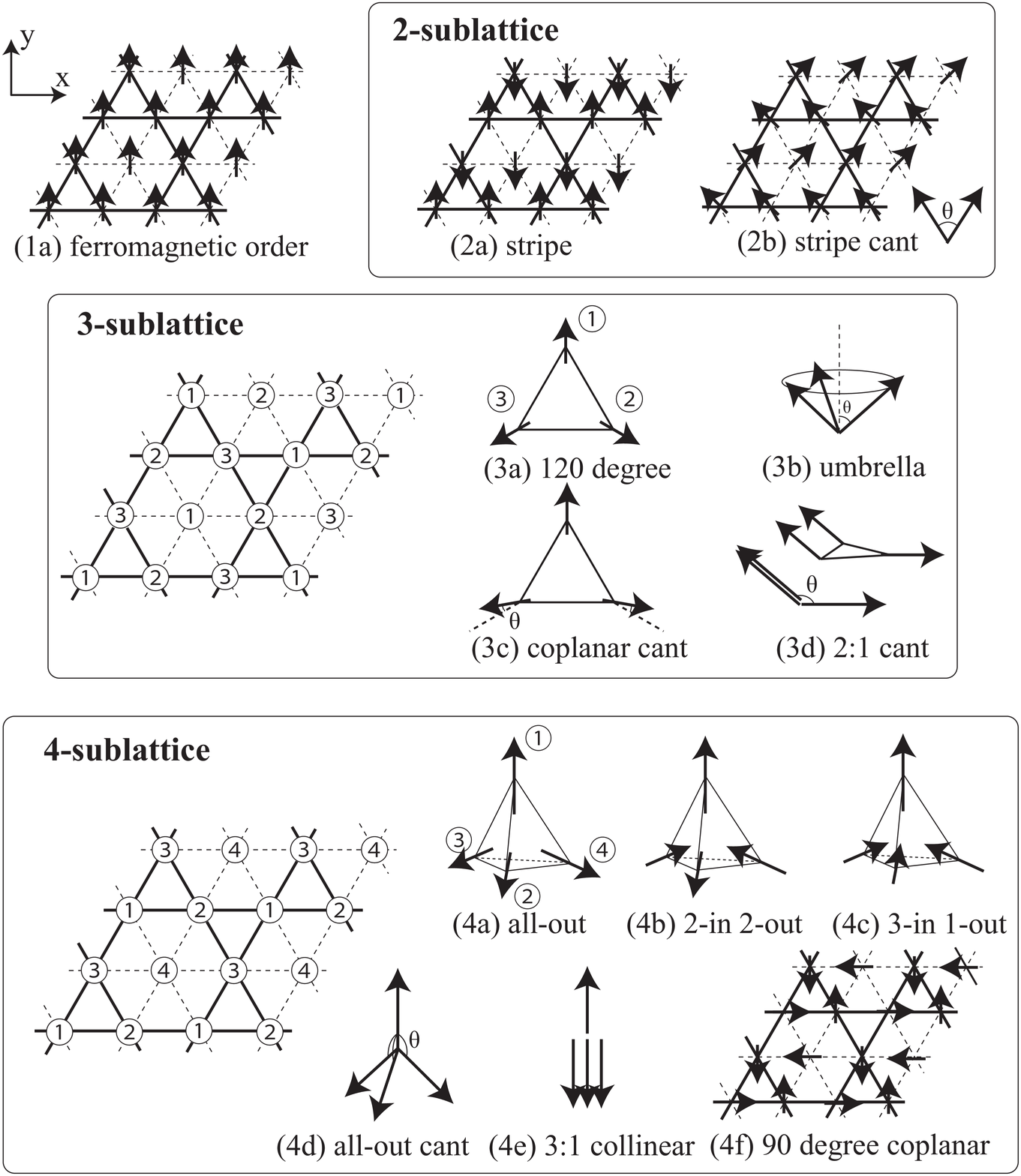}
\caption{ Ordering patterns on the triangular-to-kagome lattice: 
(1a) a ferromagnetic,
(2a) a two-sublattice collinear stripe, (2b) a stripe order with a canting angle $\theta $,
(3a) a three-sublattice $120^{\circ }$ 
noncollinear, (3b) a noncoplanar umbrella-type order with angle $\theta $ (canted
in the normal direction to the coplanar plane from the $120^{\circ }$ 
order), (3c) a coplanar order with a canting angle $\theta $ for two
spins from $120^{\circ }$  order, and (3d) a 2:1-type order with two
parallel spins that have an angle $\theta $ to the remaining one.
(4a) a four-sublattice all-out-type, (4b) a two-in
two-out-type, 
(4c) a three-in one-out-type order, (4d)
an all-out-type order with a canting angle $\theta $ for three spins,
(4e) a 3:1 collinear order,
and (4f) four-sublattice coplanar $90^{\circ }$ spiral order.
Note that (2b) with $\theta =\pi $, (3b) with $\theta =\frac{\pi }{2}$, (3c) with $\theta =0$, 
(4d) with $\theta = \cos^{-1}(-1/3)$, $\theta = \cos^{-1}(+1/3)$, and $\theta = \pi$ are equivalent to (2a), (3a), (3a), (4a), (4c), and (4e), respectively.
}\label{fig2}
\end{figure}

Following the previous studies by the authors~\cite{Akagi_2010,Akagi_2011Preprint}, 
we investigate the ground state of the model given by eq.~(1) by a variational calculation 
while varying 
the electron density $n=\frac{1}{N}\sum_{i\alpha }\langle c^{\dagger}_{i,\alpha } c_{i,\alpha }\rangle$ (\textit{N} is the total number of sites), $J_{\rm H}$, 
and $t'$. 
We compare the grand-canonical potentials $\Omega = \langle {\cal H} \rangle - \mu n$ at $T=0$ 
($\mu $ is the chemical potential) 
for different ordered states of the localized spins, and determine the most stable ordering. 
In the calculation, we consider 13 different types of ordered states, up to a four-site 
unit cell, as shown in Fig. \ref{fig2}. 
For the states (2b), (3b), (3c), (3d), and (4d), we optimize the canting angle $\theta$.

\section{Results and discussion}
\begin{figure}
\includegraphics[width=14.0cm]{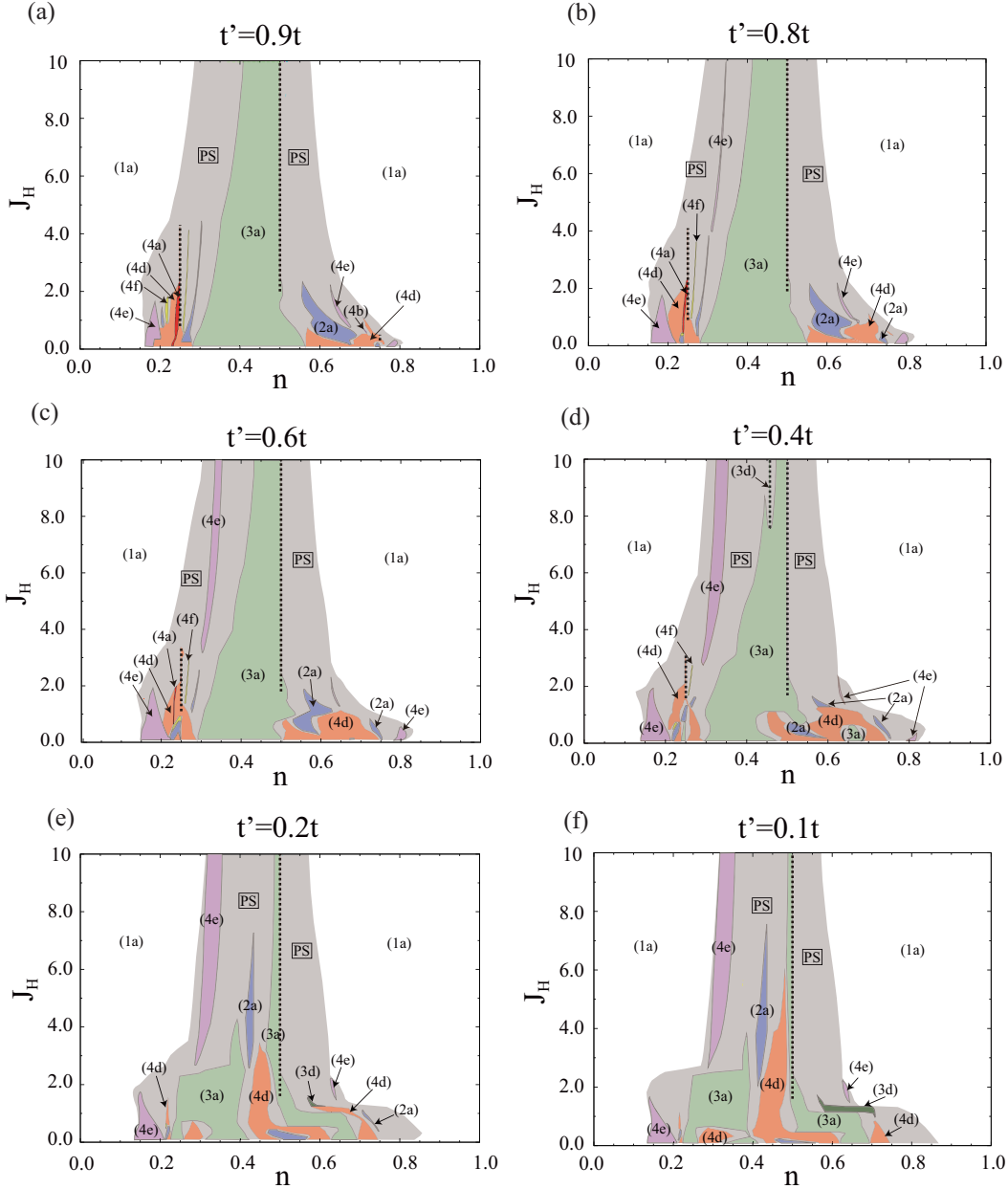}
\caption{ (Color online) Ground-state phase diagrams on the 
triangular-to-kagome lattices at 
(a) $t' = 0.9$, (b) $0.8$, (c) $0.6$,
(d) $0.4$, (e) $0.2$, and (f) $0.1$.
The vertical dashed lines at $n=1/4$, $1/2$, and $3/4$ show gapful insulating regions~\cite{comment1}. PS indicates a phase-separated region. 
}
\label{fig3}
\end{figure}

Figure~\ref{fig3} shows the results of the phase diagram as functions of $n$ and $J_{\rm H}$ at (a) $t' = 0.9$, (b) $ 0.8$, 
(c) $ 0.6 $, (d) $ 0.4$, (e) $ 0.2$, and (f) $ 0.1$.
Here, we focus on the changes of the four-sublattice chiral phases (4a) and (4d), 
which are candidates of the chiral phase with the $\mathbf{q}=\mathbf{0}$ three-sublattice order in the kagome limit $t'=0$. 
In the isotropic triangular lattice case ($t'=t=1$), 
the state (4a) appears dominantly near 1/4- and 3/4-filling, while the canted state (4d) is not stable~\cite{Akagi_2010}.
The 1/4-filling one is stabilized in a wider range of parameters than the 3/4-filling one. 

First, we focus on the change of the state (4a) near 1/4-filling. 
As we introduce a small modulation to the triangular lattice by decreasing $t'$ 
from $t'=1$, the (4a) region becomes narrower, and 
instead, the state (4d) is induced around (4a). 
When $t'$ is further decreased,
the state (4a) is taken over by (4d) around $t'\simeq 0.5$; (4d) region, however, is also narrowed. 
Eventually, the chiral states vanish in the limit of the kagome lattice ($t'=0$).
We note that the insulating state at $n=1/4$ remains robust down to $t' \simeq 0.3$, 
in sharp contrast to the 3/4-filling one described below,  
reflecting the difference of the origins of 1/4- and 3/4-filling states~\cite{Akagi_2012}.

Next, we consider the change near 3/4 filling.
When we reduce $t'$ from $t'=1$, the chiral 
state (4a) at 3/4 filling is quickly destabilized, in sharp contrast to the 1/4-filling case. 
This is because the state (4a) is stabilized by the perfect nesting of the Fermi surface 
present at $t=t'$~\cite{Martin_2008}, which disappears for $t'<t$. 
Instead, the state (4d) appears in the region of $n<3/4$. 
As decreasing $t'$, (4d) is intervened by the state (3a) 
appearing for $t'\hspace{0.3em}\raisebox{0.4ex}{$<$}\hspace{-0.75em}\raisebox{-.7ex}{$\sim $}\hspace{0.3em}0.4$.
It is noteworthy that the state (4d) spreads as $t'$ decreases, and 
is stabilized in a wide range of parameters near 1/2 filling 
for $t'\hspace{0.3em}\raisebox{0.4ex}{$<$}\hspace{-0.75em}\raisebox{-.7ex}{$\sim $}\hspace{0.3em}0.2$. 
Although this noncoplanar phase shows a finite scalar chirality and 
the associated anomalous Hall effect, 
the scalar chirality defined on the kagome network approaches 
zero as $t' \to 0$ because the canting angle $\theta $ approaches $\pi /2$. 
In the limit of the kagome lattice $t'=0$, 
the spins on the kagome network become coplanar with forming 
the ${\bf q}={\bf 0}$ $120^{\circ }$ order, 
and hence, the scalar chirality vanishes.

Consequently, our results indicate 
that the chiral phase considered by Ohgushi {\it et al.}~\cite{Ohgushi_2000} 
is not realized on a kagome lattice within the present model, 
although a related chiral phase is induced by introducing the connection to the triangular lattice by a nonzero $t'$. 
The present results, however, do not deny the possibility of another form 
of a chiral state or noncoplanar ordering on the kagome lattice, since the present calculation is 
limited by the unit cell sizes considered in the variational states in Fig.~\ref{fig2}. 
For example, it was recently pointed out that a noncoplanar order with a larger unit cell 
can be stabilized by the perfect nesting at 1/3 filling~\cite{Yu_Li_2012}.
To extend our analysis by taking account of larger unit cells is left for 
future study.

\section{CONCLUSION}
We have investigated the ground-state phase diagram of 
the Kondo lattice model with classical localized moments on triangular-to-kagome lattices.
Using the variational calculations upon a variety of ordered states up to a four-site unit cell, 
we identified the parameter regions where the spin scalar 
chiral states with noncoplanar spin configuration are stabilized. 
The chiral phases, originating in those near 1/4 and 3/4 fillings 
in the isotropic triangular lattice case, remain in a wide range of parameters 
as the lattice structure is modulated continuously to the kagome lattice. 
In the limit of kagome lattice, $t'=0$, 
however, the spin configurations
become coplanar on the kagome network 
and the scalar chirality 
vanishes in the entire region of the phase diagram. 

\begin{acknowledgments}
We acknowledge helpful discussions with Takahiro
Misawa, Masafumi Udagawa, and Youhei Yamaji. Y.A.
gratefully thanks Satoru Hayami for his fruitful comments.
Y.A. is supported by Grant-in-Aid for JSPS Fellows.
This work was supported by
Grants-in-Aid for Scientific Research (Grants
No. 19052008, No. 21340090, and No. 24340076),
Global COE Program ``the Physical Sciences Frontier",
the Strategic Programs for Innovative Research (SPIRE),
MEXT, and the Computational Materials Science Initiative
(CMSI), Japan.
The computation in this work has been done using the facilities of the
Supercomputer Center, Institute for Solid State Physics, University of Tokyo.
\end{acknowledgments}


\begin{references}
\bibitem{comment1} In addition to these states, we also find an insulating state 
(3d) with $\theta =\pi$ collinear spin configuration at a commensurate filling $n=11/24$ in Fig.~\ref{fig3}(d).
\end{references}
\end{document}